\RequirePackage{lineno}
\documentclass[aps,prl,twocolumn,showpacs,superscriptaddress]{revtex4}  

\usepackage{graphicx}
\usepackage{dcolumn}
\usepackage{bm}
\usepackage{color}
\usepackage{amssymb}
\usepackage{epstopdf}

\def\met{\mathbin{E\mkern - 11mu/_T}}
\newcommand{\unit}[1]{\,\mathrm{#1}}  

\begin{document}


\title{\boldmath Search for a New Heavy Gauge Boson $W^{\prime}$ with Electron+$\met$ Event Signature in $p\bar{p}$ collisions at $\sqrt{s}=1.96\unit{TeV}$ \unboldmath}

\affiliation{Institute of Physics, Academia Sinica, Taipei, Taiwan 11529, Republic of China} 
\affiliation{Argonne National Laboratory, Argonne, Illinois 60439, USA} 
\affiliation{University of Athens, 157 71 Athens, Greece} 
\affiliation{Institut de Fisica d'Altes Energies, Universitat Autonoma de Barcelona, E-08193, Bellaterra (Barcelona), Spain} 
\affiliation{Baylor University, Waco, Texas 76798, USA} 
\affiliation{Istituto Nazionale di Fisica Nucleare Bologna, $^z$University of Bologna, I-40127 Bologna, Italy} 
\affiliation{University of California, Davis, Davis, California 95616, USA} 
\affiliation{University of California, Los Angeles, Los Angeles, California 90024, USA} 
\affiliation{Instituto de Fisica de Cantabria, CSIC-University of Cantabria, 39005 Santander, Spain} 
\affiliation{Carnegie Mellon University, Pittsburgh, Pennsylvania 15213, USA} 
\affiliation{Enrico Fermi Institute, University of Chicago, Chicago, Illinois 60637, USA}
\affiliation{Comenius University, 842 48 Bratislava, Slovakia; Institute of Experimental Physics, 040 01 Kosice, Slovakia} 
\affiliation{Joint Institute for Nuclear Research, RU-141980 Dubna, Russia} 
\affiliation{Duke University, Durham, North Carolina 27708, USA} 
\affiliation{Fermi National Accelerator Laboratory, Batavia, Illinois 60510, USA} 
\affiliation{University of Florida, Gainesville, Florida 32611, USA} 
\affiliation{Laboratori Nazionali di Frascati, Istituto Nazionale di Fisica Nucleare, I-00044 Frascati, Italy} 
\affiliation{University of Geneva, CH-1211 Geneva 4, Switzerland} 
\affiliation{Glasgow University, Glasgow G12 8QQ, United Kingdom} 
\affiliation{Harvard University, Cambridge, Massachusetts 02138, USA} 
\affiliation{Division of High Energy Physics, Department of Physics, University of Helsinki and Helsinki Institute of Physics, FIN-00014, Helsinki, Finland} 
\affiliation{University of Illinois, Urbana, Illinois 61801, USA} 
\affiliation{The Johns Hopkins University, Baltimore, Maryland 21218, USA} 
\affiliation{Institut f\"{u}r Experimentelle Kernphysik, Karlsruhe Institute of Technology, D-76131 Karlsruhe, Germany} 
\affiliation{Center for High Energy Physics: Kyungpook National University, Daegu 702-701, Korea; Seoul National University, Seoul 151-742, Korea; Sungkyunkwan University, Suwon 440-746, Korea; Korea Institute of Science and Technology Information, Daejeon 305-806, Korea; Chonnam National University, Gwangju 500-757, Korea; Chonbuk National University, Jeonju 561-756, Korea} 
\affiliation{Ernest Orlando Lawrence Berkeley National Laboratory, Berkeley, California 94720, USA} 
\affiliation{University of Liverpool, Liverpool L69 7ZE, United Kingdom} 
\affiliation{University College London, London WC1E 6BT, United Kingdom} 
\affiliation{Centro de Investigaciones Energeticas Medioambientales y Tecnologicas, E-28040 Madrid, Spain} 
\affiliation{Massachusetts Institute of Technology, Cambridge, Massachusetts 02139, USA} 
\affiliation{Institute of Particle Physics: McGill University, Montr\'{e}al, Qu\'{e}bec, Canada H3A~2T8; Simon Fraser University, Burnaby, British Columbia, Canada V5A~1S6; University of Toronto, Toronto, Ontario, Canada M5S~1A7; and TRIUMF, Vancouver, British Columbia, Canada V6T~2A3} 
\affiliation{University of Michigan, Ann Arbor, Michigan 48109, USA} 
\affiliation{Michigan State University, East Lansing, Michigan 48824, USA}
\affiliation{Institution for Theoretical and Experimental Physics, ITEP, Moscow 117259, Russia}
\affiliation{University of New Mexico, Albuquerque, New Mexico 87131, USA} 
\affiliation{Northwestern University, Evanston, Illinois 60208, USA} 
\affiliation{The Ohio State University, Columbus, Ohio 43210, USA} 
\affiliation{Okayama University, Okayama 700-8530, Japan} 
\affiliation{Osaka City University, Osaka 588, Japan} 
\affiliation{University of Oxford, Oxford OX1 3RH, United Kingdom} 
\affiliation{Istituto Nazionale di Fisica Nucleare, Sezione di Padova-Trento, $^{aa}$University of Padova, I-35131 Padova, Italy} 
\affiliation{LPNHE, Universite Pierre et Marie Curie/IN2P3-CNRS, UMR7585, Paris, F-75252 France} 
\affiliation{University of Pennsylvania, Philadelphia, Pennsylvania 19104, USA}
\affiliation{Istituto Nazionale di Fisica Nucleare Pisa, $^{bb}$University of Pisa, $^{cc}$University of Siena and $^{dd}$Scuola Normale Superiore, I-56127 Pisa, Italy} 
\affiliation{University of Pittsburgh, Pittsburgh, Pennsylvania 15260, USA} 
\affiliation{Purdue University, West Lafayette, Indiana 47907, USA} 
\affiliation{University of Rochester, Rochester, New York 14627, USA} 
\affiliation{The Rockefeller University, New York, New York 10065, USA} 
\affiliation{Istituto Nazionale di Fisica Nucleare, Sezione di Roma 1, $^{ee}$Sapienza Universit\`{a} di Roma, I-00185 Roma, Italy} 

\affiliation{Rutgers University, Piscataway, New Jersey 08855, USA} 
\affiliation{Texas A\&M University, College Station, Texas 77843, USA} 
\affiliation{Istituto Nazionale di Fisica Nucleare Trieste/Udine, I-34100 Trieste, $^{ff}$University of Trieste/Udine, I-33100 Udine, Italy} 
\affiliation{University of Tsukuba, Tsukuba, Ibaraki 305, Japan} 
\affiliation{Tufts University, Medford, Massachusetts 02155, USA} 
\affiliation{University of Virginia, Charlottesville, VA  22906, USA}
\affiliation{Waseda University, Tokyo 169, Japan} 
\affiliation{Wayne State University, Detroit, Michigan 48201, USA} 
\affiliation{University of Wisconsin, Madison, Wisconsin 53706, USA} 
\affiliation{Yale University, New Haven, Connecticut 06520, USA} 
\author{T.~Aaltonen}
\affiliation{Division of High Energy Physics, Department of Physics, University of Helsinki and Helsinki Institute of Physics, FIN-00014, Helsinki, Finland}
\author{B.~\'{A}lvarez~Gonz\'{a}lez$^v$}
\affiliation{Instituto de Fisica de Cantabria, CSIC-University of Cantabria, 39005 Santander, Spain}
\author{S.~Amerio}
\affiliation{Istituto Nazionale di Fisica Nucleare, Sezione di Padova-Trento, $^{aa}$University of Padova, I-35131 Padova, Italy} 

\author{D.~Amidei}
\affiliation{University of Michigan, Ann Arbor, Michigan 48109, USA}
\author{A.~Anastassov}
\affiliation{Northwestern University, Evanston, Illinois 60208, USA}
\author{A.~Annovi}
\affiliation{Laboratori Nazionali di Frascati, Istituto Nazionale di Fisica Nucleare, I-00044 Frascati, Italy}
\author{J.~Antos}
\affiliation{Comenius University, 842 48 Bratislava, Slovakia; Institute of Experimental Physics, 040 01 Kosice, Slovakia}
\author{G.~Apollinari}
\affiliation{Fermi National Accelerator Laboratory, Batavia, Illinois 60510, USA}
\author{J.A.~Appel}
\affiliation{Fermi National Accelerator Laboratory, Batavia, Illinois 60510, USA}
\author{A.~Apresyan}
\affiliation{Purdue University, West Lafayette, Indiana 47907, USA}
% ...
\collaboration{CDF Collaboration\footnote{With visitors from $^a$University of Massachusetts Amherst, Amherst, Massachusetts 01003,
$^b$Istituto Nazionale di Fisica Nucleare, Sezione di Cagliari, 09042 Monserrato (Cagliari), Italy,
$^c$University of California Irvine, Irvine, CA  92697, 
$^d$University of California Santa Barbara, Santa Barbara, CA 93106
$^e$University of California Santa Cruz, Santa Cruz, CA  95064,
$^f$CERN,CH-1211 Geneva, Switzerland,
$^g$Cornell University, Ithaca, NY  14853, 
$^h$University of Cyprus, Nicosia CY-1678, Cyprus, 
$^i$University College Dublin, Dublin 4, Ireland,
$^j$University of Fukui, Fukui City, Fukui Prefecture, Japan 910-0017,
$^k$Universidad Iberoamericana, Mexico D.F., Mexico,
$^l$Iowa State University, Ames, IA  50011,
$^m$University of Iowa, Iowa City, IA  52242,
$^n$Kinki University, Higashi-Osaka City, Japan 577-8502,
$^o$Kansas State University, Manhattan, KS 66506,
$^p$University of Manchester, Manchester M13 9PL, England,
$^q$Queen Mary, University of London, London, E1 4NS, England,
$^r$Muons, Inc., Batavia, IL 60510,
$^s$Nagasaki Institute of Applied Science, Nagasaki, Japan, 
$^t$National Research Nuclear University, Moscow, Russia,
$^u$University of Notre Dame, Notre Dame, IN 46556,
$^v$Universidad de Oviedo, E-33007 Oviedo, Spain, 
$^w$Texas Tech University, Lubbock, TX  79609, 
$^x$Universidad Tecnica Federico Santa Maria, 110v Valparaiso, Chile,
$^y$Yarmouk University, Irbid 211-63, Jordan,
$^{gg}$On leave from J.~Stefan Institute, Ljubljana, Slovenia, 
}}
\noaffiliation

\begin{abstract}
We present a search for a new heavy charged vector boson $W^\prime$ decaying
to an electron-neutrino pair in $p\bar{p}$ collisions at a
center-of-mass energy of $1.96\unit{TeV}$.
The data were collected with the CDF II detector and correspond to an integrated
luminosity of $5.3\unit{fb}^{-1}$. No significant excess above the standard model
expectation is observed and we set upper limits on $\sigma\cdot{\cal B}(W^\prime\to e\nu)$.
Assuming standard model couplings to fermions and the neutrino from the $W^\prime$ boson decay to be light, 
we exclude a $W^\prime$ boson with mass less than $1.12\unit{TeV/}c^2$ at the $95\unit{\%}$ confidence level.
\end{abstract}


\maketitle

The $W^\prime$~\cite{genericWprime} is a postulated charged heavy vector boson
which is predicted in models that extend the gauge structure of the standard model.
In the left-right (LR) symmetric model~\cite{left_right} considered here, the right-handed $W^\prime$ boson mass is obtained
by the symmetry breaking of the right-handed electroweak gauge group of
$SU(2)_R \times SU(2)_L \times U(1)_{B,L}$. This provides a natural explanation for the observed suppression of $V+A$ currents in low energy weak processes. The LR symmetric model can also be motivated 
by the manifestation of a higher symmetry predicted at intermediate energies in grand unified theories~\cite{higher_symm}.

The manifest LR symmetric model assumes that the right-handed Cabibbo-Kobayashi-Maskawa matrix and the gauge coupling constants are
identical to those of the standard model~\cite{SMCKM}. The $W^\prime$ can decay in the same way as the 
standard model $W$, with the exception that the $tb$~\cite{charge conjugate} decay channel is accessible if the $W^{\prime}$ is heavy enough and that the diboson decay channel ($W^{\prime} \rightarrow WZ)$ is suppressed in the extended gauge model~\cite{genericWprime}.

The $W^\prime$ boson has been previously searched for in high energy physics experiments using final state signatures such as  
leptons, jets, and/or missing energy.
The most recent direct searches for a charged heavy vector boson have been performed at the Tevatron collider at Fermilab.
The CDF experiment previously set limits on the cross section times branching fraction
in the decay mode $W^\prime\to tb$ and excluded a $W^\prime$ boson mass below $800\unit{GeV/}c^{2}$ at the 95\% confidence level (CL) using $1.9\unit{fb}^{-1}$ data of $p\bar{p}$ collisions~\cite{cdf_wprime_tbbar}.
The D0 experiment set limits on the product of the cross section and branching fraction in the decay mode $W^\prime\to e\nu$
and excluded a $W^\prime$ boson mass
below $1.00\unit{TeV/}c^{2}$ at the 95\% CL using $1.0\unit{fb}^{-1}$ of data~\cite{d0_wprime_enu}. 
Both of these recent mass limits assume that the couplings between the new vector boson and the fermionic final states are the same as in the standard model.

In this Letter, we present the results of a search for a $W^\prime$ boson in the $e\nu$ decay mode,
assuming the manifest LR symmetric model and the right-handed neutrino from the boson decay to be light ($m_{\nu} \ll m_{W^{\prime}}$) and stable.
Under these assumptions, the results in this Letter can be useful in the generic model~\cite{genericWprime} since the kinematics of the left- and right-handed $W^\prime$ bosons is not different. 
We use a data sample corresponding to $5.3\unit{fb}^{-1}$ integrated luminosity of $p\bar{p}$ collisions
at $\sqrt{s}=1.96\unit{TeV}$ recorded by the upgraded Collider Detector at Fermilab (CDF II).
We select events that are consistent with the production of the standard model $W$ 
and the heavier $W^\prime$ boson that decay to an electron and neutrino in the final state.
The analysis technique applied is the same as in a previous search~\cite{wprime_enu205}.

The CDF II detector is described in detail elsewhere~\cite{CDF_General}.
CDF II is a general purpose solenoidal detector which combines precision charged particle tracking
with fast projective calorimetry and fine-grained muon detection.
Tracking systems are contained inside a superconducting solenoid, $1.5\unit{m}$ in radius and $4.8\unit{m}$ in length,
which generates a $1.4\unit{T}$ magnetic field parallel to the beam axis.
Calorimeters and muon systems surround the solenoid and the tracking system.
Electron candidates are identified by an energy deposit 
in the electromagnetic calorimeter with a track pointing to it.
A set of charged-particle detectors surrounding the calorimeters identify muon candidates.
The energy of the electron candidate is measured by the calorimeter and 
its direction is determined from the tracking system.
The component of the neutrino momentum transverse to the beamline is inferred to be equal to the missing
transverse energy $\met$~\cite{met}, which is derived from the transverse energy imbalance of all the deposited energy in the calorimeters.

The online selection requires either one electron candidate in the electromagnetic calorimeter
with transverse energy $E_T>18\unit{GeV}$ that has a matching track with transverse momentum $p_T>9\unit{GeV}/c$ or an
electron candidate in the electromagnetic calorimeter with transverse energy $E_T>70\unit{GeV}$. 
No restrictions on the amount of energy leakage into the hadronic calorimeter
were imposed, in order to ensure high efficiency for high-$E_T$ electrons.
We select the candidate event sample offline
by requiring an isolated electron candidate with
$E_T>25\unit{GeV}$ and the existence of an associated track with $p_T > 15\unit{GeV}/c$ that is contained in the fiducial region 
of the tracking system of $|\eta|<1.0$~\cite{coordinate system}.
Electron candidates are selected based on an $E_T$-dependent isolation cut~\cite{isolation} in order to
maximize the efficiency in the high-$E_T$ region.
The electron shower profile is required to be consistent with that of test-beam electrons in order to match with the expected EM shower~\cite{shower profile}.
In events with high-energy muons, the $\met$ is adjusted by adding the muon momentum and
removing the expected ionization energy deposition in the calorimeter.
The $\met$ is corrected further for $\eta$- and energy-dependent non-uniformities
of the calorimeter response. In the final selection, the corrected $\met$ is required to be greater than $25\unit{GeV}$.
Dilepton events coming from Drell-Yan, $t\bar{t}$, and diboson backgrounds are vetoed
by rejecting events with a second isolated lepton, either an electron or a muon, with $p_{T} > 15\unit{GeV/c}$.
QCD multijet events are a background to $W/W^{\prime}\to e\nu$ when a jet is misidentified as an electron and mismeasured jets lead to significant $\met$.
The electron candidate $E_T$ and the event $\met$ are likely to significantly differ in magnitude in this case.
In contrast, a $W/W^\prime\to e\nu$ event will have an electron and neutrino emitted in opposite directions
which results in the electron $E_T$ and $\met$ being of comparable magnitude, respectively,
assuming the $p_T$ of the boson is much smaller than its mass.
Thus, in order to reduce the QCD multijet background,
we require the candidate events to satisfy $0.4 < E_T /\met < 2.5$. 
The efficiency of this requirement is larger than $99\unit{\%}$ for $W/W^{\prime}$ events whereas the rejection rate is $\sim40\unit{\%}$ for QCD multijet events with $E_{T}>100\unit{GeV}$.
After all selection requirements, the transverse mass of a candidate event is calculated as
\begin{equation}
m_T\equiv\sqrt{2E_T\met(1-\cos\phi_{e\nu})},
\end{equation}
where $\phi_{e \nu}$ is the azimuthal opening angle between 
the electron candidate and the $\met$ direction. 
\begin{figure}[ht]
\begin{center}
\includegraphics[width=1\columnwidth]{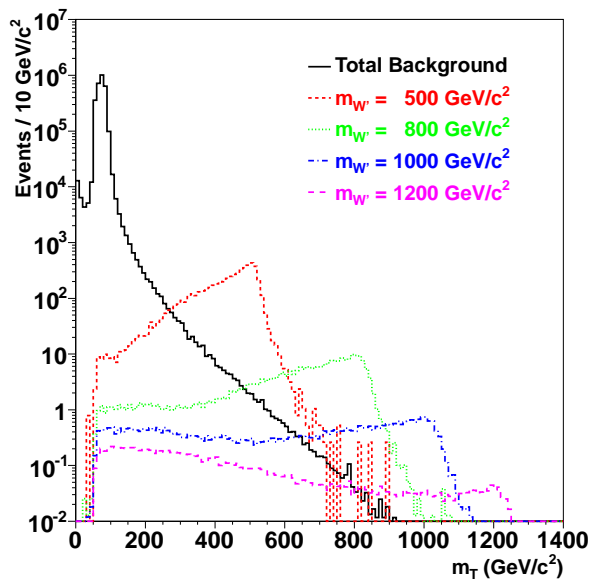}
\caption{The transverse mass distributions for $W^{\prime} \rightarrow e \nu$ signal events generated using {\scshape pythia} with total background expectation.}
\label{FigWprimeMass}
\end{center}
\end{figure}

The $W^\prime\to e\nu$ signal events are generated with {\scshape pythia}~\cite{pythia} using the CTEQ5L~\cite{cteq}
parton distribution functions (PDFs) and a simulation of the CDF II detector.
Since the cross sections calculated by {\scshape pythia} are at leading order (LO), next-to-next-to-leading-order (NNLO) K factors are applied to the LO cross sections.
Mass-dependent NNLO K factors from Ref.~\cite{hamberg} are obtained with an approximately magnitude around 1.3.
The total acceptance times efficiency of the event selection 
cuts ranges from $45\unit{\%}$ to $35\unit{\%}$ and decreases above a $W^\prime$ boson mass of $800\unit{GeV/}c^{2}$.
Figure~\ref{FigWprimeMass} shows the expected $W^{\prime}$ boson transverse mass distributions for various input masses with the background predictions.
The on-shell production of heavy bosons near the kinematic limit is suppressed due to
the smallness of the PDFs at large momentum fraction, which results in the low acceptance rate of $W^\prime$ events at high mass above $800\unit{GeV/}c^{2}$ after applying the kinematic selection requirements.

\begin{figure}[ht]
\begin{center}
\includegraphics[width=1\columnwidth]{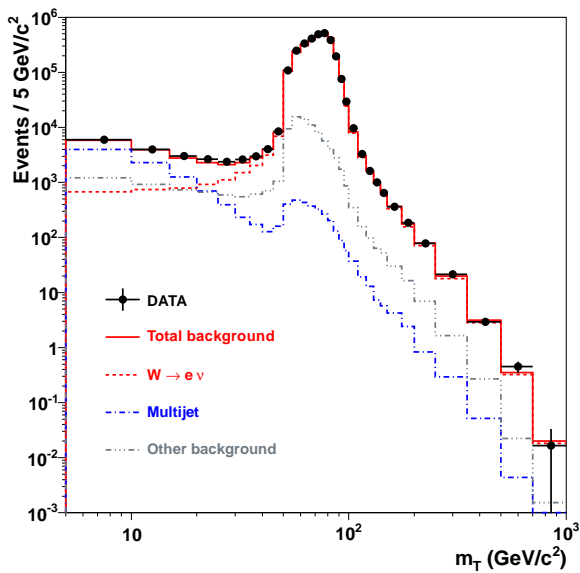}
\caption{The transverse mass distributions of $e\nu$ candidate events compared to the total backgrounds.}
\label{fig:mtfit}
\end{center}
\end{figure}

The background sources to $W^\prime\to e\nu$ are primarily processes with an 
electron and missing energy in the final state.
These sources of background are $W\to e\nu$, $W\to \tau\nu\to e\nu\nu\nu$, $Z/\gamma^*\to\tau\tau\to eX$, $tt$, and diboson ($WW$, $WZ$) production.
The $Z/\gamma^*\to ee$ process can also produce missing energy
when one of the electrons escapes detection.
The $m_T$ distributions and acceptance times efficiency of the non-multijet backgrounds
are obtained using {\scshape pythia} and a simulation of the CDF II detector.
Theoretical cross section predictions are used to estimate the expected background yields~\cite{hamberg,xsecDiboson,xsecTTbar}.
For the QCD multijet background estimation, a data-driven method is applied that uses the distribution of the azimuthal angle between 
the primary electron candidate and the vector sum of the jet energy.
For the multijet case, a jet misidentified as an electron candidate will appear to recoil against the rest of the jet in the event. 
Therefore, a back-to-back distribution is expected in the azimuthal opening angle. The $W/W^{\prime} \rightarrow e\nu$ process however does not have a strong correlation in this angle. The QCD multijet contribution is estimated by a likelihood fit to the data using the different angular shapes.
The multijet $m_{T}$ distribution is obtained using a QCD enriched sideband sample with the isolation cut inverted.
The data and the total background $m_T$ distributions are compared in Fig.~\ref{fig:mtfit}.
The contributions from $W\to e\nu$, QCD multijet, 
and the other backgrounds in the mass region above
$m_T=200\unit{GeV/}c^{2}$ are listed in Table~\ref{tab:nfit_bin}.
This comparison shows good agreement between the data and the total backgrounds.

\begin{table*}[ht]
\begin{center}
\caption{The event yields for the background sources in $m_{T}$ above $200\unit{GeV/c^{2}}$ compared to the observed data.}
\def\arraystretch{1.5}
\begin{tabular}{c|ccccc}
\hline\hline
   &\multicolumn{5}{c}{Events in $m_T$ bins ($\unit{GeV/c^{2}}$) }\\
   &200 - 250 &250 - 350 &350 - 500 &500 - 700 &700 - 1000   \\\hline
$W \to e\nu$      & $711_{-50}^{+50}$&  $359_{-25}^{+25}$ &  $85_{-6}^{+6}$ &  $13_{-1}^{+1}$     & $1.1_{-0.1}^{+0.1}$ \\
Multijet          & $  9_{-2}^{+2}$  &  $  6_{-1}^{+1}$   &  $ 2_{-2}^{+2}$ &  $0.2_{-0.2}^{+1.6}$& $0.01_{-0.01}^{+1.10}$ \\
Other background  & $ 70_{-6}^{+9}$  &  $ 33_{-3}^{+4}$   &  $ 8_{-1}^{+1}$ &  $ 1_{-0.1}^{+0.1}$ & $0.09_{-0.01}^{+0.01}$ \\ \hline
Total background  & $790_{-58}^{+61}$&  $398_{-30}^{+31}$ &  $94_{-8}^{+9}$ &  $14_{-1}^{+3}$     & $1.2_{-0.1}^{+1.2}$ \\ \hline
Data              & $784            $&  $426            $ &  $88          $ &  $18              $ & $1            $ \\
\hline \hline
\end{tabular}
\end{center}
\label{tab:nfit_bin}
\end{table*}

In order to quantify the size of the potential signal contributions
in the data sample, a binned maximum likelihood fit was performed on the
observed $m_T$ distribution between $0$ and $1500\unit{GeV/c^2}$,
using the background predictions and the expected $W^\prime$ boson contribution
for different mass values ranging from $500$ to $1300\unit{GeV/c^2}$.
The fit results are shown in Table~\ref{tab:valuefit}, normalized to
\begin{equation}
\beta \equiv {\sigma\cdot{\cal B}(W^\prime\to e\nu)\over
\sigma\cdot{\cal B}(W^\prime\to e\nu)_{LR}},
\end{equation}
where the numerator is the observed cross section times branching fraction 
and the denominator is that expected from the manifest LR symmetric model.
The expected signal yield was normalized to the observed $W$ boson yield obtained from the fit.
This removes several sources of systematic uncertainty such as the integrated luminosity, 
the trigger and the identification efficiencies, all of which cancel in the ratio.

Systematic uncertainties 
on the signal and the background rates 
were considered for the PDFs, the jet energy scale,
the theoretical cross sections, the multijet background, the initial/final state radiation of the signal,
and the energy scale of the electromagnetic calorimeter. 
The dominant contribution to the systematic uncertainty
comes from the PDFs.
The total systematic
uncertainty varies from $\pm5\unit{\%}$ to $\pm10\unit{\%}$ for $W^\prime$ boson masses ranging from
$m_{W^\prime}=500$ to $1300\unit{GeV/c^2}$.

\begin{figure}[ht]
\begin{center}
\includegraphics[width=1\columnwidth]{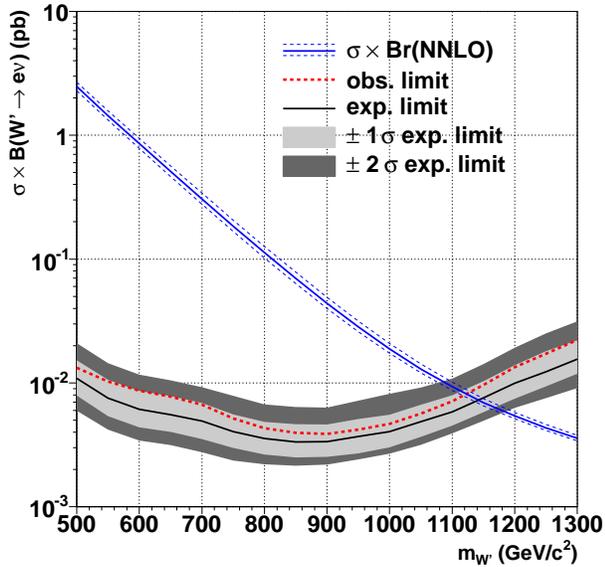}
\caption{The $95\unit{\%}$ CL limits on cross section times branching fraction
as a function of $W^\prime$ boson mass and the expected limits from the simulated experiments with background only.
The region above the red dashed line is excluded at the $95\unit{\%}$ CL. 
The cross section times branching fraction 
assuming the manifest LR symmetric model,
$\sigma\cdot{\cal B}(W^\prime\to e\nu)_{LR}$, is shown along with
its uncertainty. The intercept of the cross section limit curve and
the lower bound of the theoretical cross section yields
$m_{W^\prime}>1.12\unit{TeV/c^2}$ at the $95\unit{\%}$ CL.}
\label{fig:limits}
\end{center}
\end{figure}

To determine the limit on $\beta$, we use a Bayesian approach~\cite{mclimit} by constructing a marginalized posterior probability distribution ($p(\beta)$) from the likelihood function.
Sources of systematic uncertainty are included as nuisance parameters in the definition of the likelihood function.
The $95\unit{\%}$ CL upper limits on the ratio of the observed to the expected cross section
are obtained from the fit.
We use the resulting likelihood function only in the ``physical region'' where this ratio is greater than or equal to zero.
The obtained upper limits are summarized in Table~\ref{tab:valuefit} and
plotted in Figure~\ref{fig:limits} as a function $m_{W^\prime}$ together with the expected limits obtained from simulated experiments with background only.
Using theoretical predictions that assume the manifest LR symmetric model~\cite{SMCKM},
the limits on the cross section times branching
fraction are converted into limits on the mass of the $W^\prime$ boson.
The lower mass limit can be set at the mass value for which $\beta_{95} = 1$, where $\int_{0}^{\beta_{95}}p(\beta)d\beta = 0.95$.
We take the lower bound of the theoretical cross section to obtain the mass limit.
Hence, the $95\unit{\%}$ CL is found to be $m_{W^{\prime}} > 1.12\unit{TeV/c^2}$.
\begin{table}[ht]
\begin{center}
\caption{The expected numbers of events from $W^\prime\to e\nu$ process,
$N_{\rm{exp}}$, assuming the manifest LR symmetric model and normalized by
the observed $W$ boson yield.
We also show the observed relative rate of the $W^\prime$ boson production
from the fit described in the text,
and the 95\% CL upper limit on this relative rate.
The uncertainties are statistical only and do not include systematic uncertainties.
The 95\% upper limits include both statistical and systematic uncertainties.}
\def\arraystretch{1.0}
\begin{tabular}{cccc}
\hline\hline
$m_{W^\prime}$   &$N_{\rm{exp}}$
&\multicolumn{2}{c}{$\beta \left(={\sigma\cdot{\cal B}(W^\prime\to e\nu)\over
\sigma\cdot{\cal B}(W^\prime\to e\nu)_{LR}}\right)$} \\
(GeV/c$^{2})$          &(events)            &Fit ($\times 10^{-2}$)   &Upper Limit   \\\hline
500    & 5828    & $0.08_{-0.08}^{+0.21}$ & 5.38 $\times 10^{-3}$ \\
550    & 3407    & $0.18_{-0.18}^{+0.26}$ & 7.16 $\times 10^{-3}$ \\
600    & 2037    & $0.28_{-0.28}^{+0.36}$ & 1.01 $\times 10^{-2}$ \\
650    & 1218    & $0.43_{-0.43}^{+0.54}$ & 1.52 $\times 10^{-2}$ \\
700    &  731    & $0.36_{-0.36}^{+0.83}$ & 2.22 $\times 10^{-2}$ \\
750    &  433    & $0.15_{-0.15}^{+1.07}$ & 2.80 $\times 10^{-2}$ \\
800    &  263    & $0.03_{-0.03}^{+1.36}$ & 3.82 $\times 10^{-2}$ \\
850    &  160    & $0.00_{-0.00}^{+1.89}$ & 5.68 $\times 10^{-2}$ \\
900    &  100    & $0.00_{-0.00}^{+2.80}$ & 8.79 $\times 10^{-2}$ \\
950    &   62    & $0.00_{-0.00}^{+4.53}$ & 1.49 $\times 10^{-1}$ \\
1000   &   41    & $0.00_{-0.00}^{+6.64}$ & 2.48 $\times 10^{-1}$ \\
1050   &   27    & $0.00_{-0.00}^{+10.8}$ & 4.36 $\times 10^{-1}$ \\
1100   &   19    & $0.00_{-0.00}^{+17.7}$ & 7.62 $\times 10^{-1}$ \\
1150   &   14    & $0.00_{-0.00}^{+32.5}$ & 1.39                  \\
1200   &   10    & $0.00_{-0.00}^{+62.7}$ & 2.47                  \\
1250   &   8.1   & $0.00_{-0.00}^{+114}$  & 3.96                  \\
1300   &   6.7   & $0.00_{-0.00}^{+224}$  & 6.24                  \\
\hline\hline
\end{tabular}
\label{tab:valuefit}
\end{center}
\end{table}

In summary, we have performed a search for a new heavy charged vector
boson decaying to an electron-neutrino pair with a light and stable
neutrino in $p\bar p$ collisions at $\sqrt{s} = 1.96\unit{TeV}$.
We do not observe any statistically significant excess over the background
expectations. We use a fit to the $m_T$ distribution to set upper limits
on the production and decay rate of a $W^\prime$ boson as a function of $m_{W^\prime}$,
and exclude a $W^\prime$ boson with $m_{W^\prime}<1.12\unit{TeV/c^2}$ at
the $95\unit{\%}$ CL, assuming the manifest LR symmetric model.

\begin{acknowledgments}
We thank the Fermilab staff and the technical staffs of the participating institutions for their vital contributions.
This work was supported by the U.S. Department of Energy and National Science Foundation;
the Italian Istituto Nazionale di Fisica Nucleare;
the Ministry of Education, Culture, Sports, Science and Technology of Japan;
the Natural Sciences and Engineering Research Council of Canada;
the National Science Council of the Republic of China;
the Swiss National Science Foundation;
the A.P. Sloan Foundation;
the Bundesministerium f\"ur Balduin Una Forschung, Germany;
the World Class University Program, the National Research Foundation of Korea;
the Science and Technology Facilities Council and the Royal Society, UK;
the Institut National de Physique Nucleaire et Physique des Particules/CNRS and Universite Pierre et Marie Curie;
the Russian Foundation for Basic Research;
the Ministerio de Ciencia e Innovaci\'{o}n, and Programa Consolider-Ingenio 2010, Spain;
the Slovak R\&D Agency;
and the Academy of Finland.
\end{acknowledgments}


\begin{thebibliography}{99}
\bibitem{charge conjugate}
We omit charge, anti-particle, and flavor-tagging neutrino symbols when representing
the decay modes but assume proper charge conjugation throughout
this report. For example, $W^{\prime+}\to e^+\nu_{e}$ and its charge
conjugate are denoted as $W^\prime\to e\nu$.
\bibitem{met}
Missing transverse energy, $\met$, is defined as the magnitude of
$-\sum_{i}E_T^i\hat{n}_i$, where $\hat{n}_i$ is a unit vector 
in the azimuthal plane that points from the beam line to the center of the $i$th calorimeter tower.
\bibitem{coordinate system}
We use a coordinate system where $\theta$ is the polar angle to the proton beam,
$\phi$ is the azimuthal angle about the beam axis, and $\eta$ is the
pseudorapidity defined as $-\ln(\tan(\theta/2))$.
Energy (track momentum) measured transverse to the beam line is denoted
as $E_T$ ($p_T$).
\bibitem{isolation}
The additional energy ($\equiv E_{T}^{0.4} - E_T^{ele}$) must be less then $3 + 0.02 \times E_{T}^{ele}$, 
where $E_{T}^{0.4}$ is the transverse energy in the cone of 
$\Delta R(\equiv \sqrt{(\Delta \phi)^{2} + (\Delta \eta)^{2}}) <0.4$ centered on the electron track and $E_{T}^{ele}$ is the transverse
energy of the electron candidate. Both energies are in units of GeV.
\end{thebibliography}
\end{document}